\documentclass{iopart}

\usepackage{graphicx,amsfonts,amsbsy,amssymb,amsthm}
\usepackage{amsopn}
\usepackage{color}
\usepackage{cite}
\usepackage{hyperref}

\DeclareMathOperator{\Var}{Var}
\DeclareMathOperator{\Res}{Res}

\eqnobysec
\newcommand{\eqref}[1]{(\ref{#1})}

\newcommand{\BN}{\mathfrak{B}_N}
\newcommand{\E}{\mathbf{E}}

\newcommand{\blambda}{\boldsymbol{\lambda}}
\newcommand{\bt}{\boldsymbol{t}}

\newcommand{\p}{\partial}
\renewcommand{\vec}[1]{| #1 \rangle}
\newcommand{\tvec}[1]{\langle #1 |}
\newcommand{\mean}[1]{\langle #1 \rangle}

\newtheorem{prop}{Proposition}
\newtheorem*{prop*}{Proposition}

\def\={\;=\;} \def\+{\,+\,} \def\m{\,-\,}    \def\p{\partial}
\def\C{\Bbb C}      
\def\l{\lambda}  \def\a{\alpha}  \def\b{\beta}  \def\D{\Delta} \def\d{\delta} \def\e{\varepsilon}

\begin{document}
 
\title{Dyson's Brownian-motion model for random matrix theory - revisited}

\author{Christopher H. Joyner$^{1,2}$  and Uzy Smilansky$^{1}$ (with an appendix by Don B. Zagier$^{3}$)}
\address{$^{1}$Department of Physics of Complex Systems, Weizmann Institute of Science, Rehovot 7610001, Israel.}
\address{$^{2}$School of Mathematical Sciences, Queen Mary University of London, London, E1 4NS, UK}
\address{$^{3}$ Max-Planck Institut f\"ur Mathematik, Vivatsgasse 7, D-53111 Bonn, Germany.}
 \ead{\mailto{c.joyner@qmul.ac.uk}, \mailto{uzy.smilansky@weizmann.ac.il},
 \mailto{dbz@mpim-bonn.mpg.de}.}

\begin{abstract}
We offer an alternative viewpoint on Dyson's original paper regarding the application of Brownian 
motion to random matrix theory (RMT). In particular we show how one may use the same approach in 
order to study the stochastic motion in the space of matrix traces $t_n = \sum_{\nu=1}^{N} \lambda_\nu^n$, 
rather than the eigenvalues $\lambda_\nu$. In complete analogy with Dyson we obtain a Fokker-Planck 
equation that exhibits a stationary solution corresponding to the joint probability density function 
in the space $\bt = (t_1,\ldots,t_n)$, which can in turn be related to the eigenvalues 
$\blambda = (\lambda_1,\ldots,\lambda_N)$. As a consequence two interesting combinatorial 
identities emerge, which are proved algebraically in the appendix.  We also offer a number 
of comments on this version of Dyson's theory and discuss its potential advantages.

\end{abstract}

\section {Introduction}
\label{introduction}
In his seminal 1962 paper, {\it A Brownian-Motion Model for the Eigenvalues of a Random Matrix} \cite{Dyson1}, 
F.~Dyson provided a conceptually novel and practical approach to the theory of random matrices, paving the 
way for many interesting developments (see e.g. \cite{mehtabook,forrester,Anderson,Guhr,Akemann,Haakebook,Porter} 
and references cited therein.) In it he explains how to introduce a dynamical approach to the theory of 
random matrices and the traditional Gaussian ensembles in particular. We briefly recapitulate the 
results here in this introductory section.

Consider a self adjoint matrix $M$ of size $N \times N$, whose entries are of the form 
$M_{i j}=\sum_{\alpha=0}^{\beta-1}M_{ij;\alpha}e_{\alpha}$. The coefficients $M_{ij;\alpha}$ being 
real parameters and $e_{\alpha}$ are the units of the three potential algebras: real ($\beta = 1)$, 
complex ($\beta=2$) and real-quaternion ($\beta=4$), satisfying $e_0^2 =1$ and $e_\alpha^2 = -1 \ \forall\  \alpha>0$. 
Choosing the real coefficients $M_{ij;\alpha}$ independently from a Gaussian distribution with zero mean and 
variance $\E(M_{ij;\alpha}^2) = (1+\delta_{ij})/(2\beta)$ we obtain the Gaussian orthogonal, unitary and 
symplectic ensembles (GOE, GUE and GSE) for $\beta=1,2$ and 4 respectively. Thus the probability distribution 
for the matrix $M$ may be neatly summarised in the following form
\begin{equation}\label{Matrix prob dist}
P(M) = \kappa_{\beta}^{(N)} \ {\rm e}^{-\frac{\beta}{2}\tr MM^\dagger},
\end{equation}
with $\kappa_{\beta}^{(N)} $ a normalization constant.

Crucially, Dyson realised that the above distribution can be identified as the stationary distribution of 
a Brownian particle in $N + \beta N(N-1)/2$ dimensions. More precisely, this means that each independent 
element $M_{ij;\alpha}, 1 \leq i \leq j\leq N$ undergoes a 1D Ornstein-Uhlenbeck process, so that in the 
(fictitious) time $s$ the motion of $M_{ij;\alpha}$ is completely determined by the following moments:
\begin{eqnarray}\label{Matrix elements general 1}
\E(\delta M_{ij;\alpha})  & = & - M_{ij,\alpha}\delta s \\
\label{Matrix elements general 2}
\E(\delta M_{ij;\alpha}^2) & = & \frac{1}{\beta}(1 + \delta_{ij})\delta s . 
\end{eqnarray}
The latter implies that $\E(|\delta M_{ij}|^2) =  (1 + (2/\beta - 1)\delta_{ij})\delta s$ (since the 
diagonal elements $M_{ii}$ are always real).

Importantly, this stochastic motion is invariant under unitary transformations, meaning the eigenvectors 
do not play any role in the corresponding motion induced in the $N$ dimensional space of eigenvalues 
$\blambda =(\lambda_1,\cdots, \lambda_N)$. Therefore one may choose a representation in which $M$ is 
diagonal, leading to a perturbation of the eigenvalue $\lambda_{\mu}$ due to a small change in the matrix $\delta M$ of
\begin{equation}\label{Perturbation formula}
\delta \lambda_{\mu} = \delta M_{\mu\mu;0} + \sum_{\nu \neq \mu} \frac{|\delta M_{\mu\nu}|^2}{\lambda_{\mu} - \lambda_{\nu}}.
\end{equation}
Obtaining the first two moments of the evolution then follows directly from the expressions
(\ref{Matrix elements general 1}) and (\ref{Matrix elements general 2}), given by
\begin{eqnarray}\label{Eigenvalue motion 1}
\E(\delta \lambda_{\mu})  & = & F_\mu(\blambda)\delta s 
= \left[\sum_{\nu\ne\mu} \frac{1}{\lambda_{\nu}-\lambda_\mu} - \lambda_{\mu}\ \right]\delta s \\
\label{Eigenvalue motion 2}
\E(\delta \lambda_{\mu}^2) & = & \frac{2}{\beta}\delta s .
\end{eqnarray}
Using these two moments, one obtains a Fokker-Planck equation that describes how the joint probability 
distribution function (JPDF) $P(\blambda;s)$ evolves in time, given some specific initial distribution 
$P(\blambda;0)$;
\begin{equation}
\label{FP-spectral}
\frac{\partial P}{\partial s}=\sum_{\mu=1}^{N}\left [- \frac{\partial }{\partial \lambda_{\mu}}(F_{\mu}(\blambda)P(\blambda;s)) 
+\beta^{-1} \frac{\partial^2P(\blambda;s)}{\partial \lambda_{\mu}^2}\right].
\end{equation}
The real advantage, and one might add elegance, of this approach is expressed in the above equation. 
In general it is not known how to obtain $P(\blambda;s)$ for arbitrary initial conditions and times $s$. 
However, since we are interested in the stationary distribution, we can reduce the complexity by setting 
the LHS equal to zero, at which point one solves the equation easily to obtain
\begin{equation}
\label{spectrJPD}
P(\blambda)=  C^{(N)}_{\beta} \prod_{\mu<\nu}|\lambda_{\mu}-\lambda_{\nu}|^{\beta} 
{\rm exp}\left(-\frac{\beta}{2} \sum_{\mu}\lambda_{\mu}^2\right ),
\end{equation}
with $C^{(N)}_{\beta}$ a normalisation constant (see e.g. Chapter 3 of \cite{mehtabook}). 
Moreover, since we know that the underlying motion (\ref{Matrix elements general 1}) and 
(\ref{Matrix elements general 2}) in the space of matrices leads to the probability distribution 
(\ref{Matrix prob dist}), the expression (\ref{spectrJPD}) must be the unique stationary distribution 
for the process (\ref{Eigenvalue motion 1}) and (\ref{Eigenvalue motion 2}) and is therefore the 
JPDF of the eigenvalues in the appropriate Gaussian ensembles.

The key component of (\ref{spectrJPD}) is the Vandermonde determinant $\prod_{\mu<\nu} |\lambda_{\mu}-\lambda_{\nu}|$, 
which is responsible for the apparent repulsion of neighbouring eigenvalues. This factor emerges as the 
Jacobian of the transformation from (\ref{Matrix prob dist}) to (\ref{spectrJPD}). However, as Dyson 
highlights, the above approach offers a new insight into its appearance - as it is nothing more than 
the effect coming from the second order term in the perturbation formula (\ref{Perturbation formula}).

Recently, the authors have adapted the above approach to investigate the spectral statistics of 
\emph{Bernoulli} matrices \cite{Joyner-2015} (matrices in which the elements come from the set $\{\pm 1\}$). 
In this instance higher terms in the perturbation formula had to be accounted for, which meant assumptions 
regarding the delocalisation of eigenvectors were required. This inevitably led to the following 
question - can Dyson's Brownian motion model be used without the requirement of the perturbation 
formula (\ref{Perturbation formula})?

In this article we demonstrate that the answer is indeed positive. To achieve this we start from a 
slightly different viewpoint to Dyson: Rather than following the evolution $P(\blambda;s)$ of the 
eigenvalues directly, we instead follow $Q(\bt;s)$ - the JPDF of the $N$-dimensional vector of traces 
$\bt = (t_1,\cdots,t_N)$, where $t_k=\sum_{\nu=1}^N\lambda_\nu^k = \tr M^k$. Performing a transformation 
of variables then allows us to recover the stationary solution (\ref{spectrJPD}) expressed in terms 
of the $\bt$ variables.

To the best of our knowledge, the distribution of the traces (or spectral moments) has  not been 
extensively studied, although there are exceptions for both the Gaussian and circular ensembles 
(see e.g. \cite{Sinai,Essen,Guionnet,Diaconis-1994,Diaconis-2004} and references therein). We therefore 
find it worthwhile to pursue this direction, not only as it sheds new light on Dyson's approach, but 
because it may offer different perspectives on such trace distributions. In addition, our method has 
led to the discovery of two identities (see Proposition \ref{Identities}) that relate the traces $t_n$ 
with $n > N$ to those with $n \leq N$. We are unaware of the existence of similar identities in the 
literature and a direct proof of their validity has kindly been supplied by D. Zagier in \ref{Zagier proof}.

It has also been brought to our attention\footnote{For which we would like to thank P. Forrester.} 
that a similar philosophy has also been undertaken by Bakry and Zani \cite{Bakry-2014}. Rather than 
looking at the traces $\bt$ they follow the motion of the secular coefficients (given by $c_k$ in 
Section~\ref{definitions}). Their motivation comes from wanting to generalise the probability density 
functions to Gaussian random matrices with Clifford algebras (rather than real, complex or quaternion 
entries) and they too note that such approaches have not been utilised before.

The paper is organised as follows: In Section \ref{definitions} we introduce the basic concepts to 
be discussed, provide some useful relations and outline the identities mentioned above. We also provide 
explicit formulae for the stationary distribution $Q_{\beta}(\bt)$ for $\beta = 1,2,4$ and arbitrary 
dimension $N$. In Section \ref{F-P equations} we derive the Fokker-Planck equation for $Q(\bt;s)$ and 
give an example of its form in two dimensions in Section \ref{2 Dim example}. In Section 
\ref{Sec: Stationary solution} we analyse the equation in $N$-dimensions and show how the aforementioned 
identities arise from considering the stationary solution $Q_{\beta}(\bt)$. Section \ref{Sec: Mean values} 
is used briefly to explain how the mean spectral moments also arise naturally in this context. Finally 
in Section \ref{Sec: Application} and Section \ref{Sec: Discussion} we provide an application of this 
method to Bernoulli ensembles and discuss the potential advantages of the whole approach.

\section {Definitions and useful relations}
\label{definitions}
The first essential feature to outline is the relationship between the spectral and trace distribution 
functions $P(\blambda;s)$ and $Q(\bt;s)$. The elements of the Jacobian of the transformation are 
given by $\frac{\partial t_n}{\partial \lambda_\nu} = n \lambda_\nu^{n-1}$, which means that
\begin{equation}\label{PD transformation}
P(\blambda;s)  = \left|\frac{\partial \bt}{\partial \blambda}\right| Q(\bt;s)  = N! \det(V) \; Q(\bt;s) .
\end{equation}
Here $V$ is the familiar Vandermonde matrix
\begin{equation}\label{Vandermonde}
V=
\left(\begin{array}{cccc}
1 & 1 & \cdots & 1 \\
\lambda_1 & \lambda_2 & \cdots & \lambda_N \\
\lambda_1^2 & \lambda_2^2 & \cdots & \lambda_N^2 \\
\vdots & \vdots & \ddots & \vdots \\
\lambda_1^{N-1} & \lambda_2^{N-1} & \cdots & \lambda_N^{N-1} \end{array}\right)
\end{equation}
and so $\det(V) = \prod_{\mu < \nu} | \lambda_{\mu} - \lambda_{\nu}|$, as seen in (\ref {spectrJPD}). 
The mapping $\blambda\rightarrow \bt$ is one-to-one as long as the Jacobian does not vanish, hence 
we must restrict the spectral variables to an ordered sector, e.g. $\lambda_1 < \lambda_2 < \cdots < \lambda_N$. 
In order to obtain an expression for $Q(\bt;s)$ we need to write $P(\blambda;s)$, and thus the Vandermonde 
determinant $\det(V)$, in terms of the traces $\bt$. Fortunately this is relatively straightforward, 
since $G(\bt) = \det(V) = \sqrt{\det(VV^{\intercal})}$, with
\[
VV^{\intercal} =
\left(\begin{array}{cccc}
t_0 & t_1 & \cdots & t_{N-1} \\
t_1 & t_2 & \cdots & t_N \\
t_2 & t_3 & \cdots & t_{N+1} \\
\vdots & \vdots & \ddots & \vdots \\
t_{N-1} & t_N & \cdots & t_{2N-2} \end{array}\right)
\]
and $t_0=N$  (see \cite{Dunne-1993,Vivo-2008} for example for uses of this identity in other contexts). 
At this point $G(\bt)$ is expressed entirely in terms of the traces, as desired, however this includes 
traces of higher degree than $N$, which are themselves functions of the traces $t_n, \ 1\le n\le N$. The  
expressions for $t_{N+r}$ in terms of the first $N$ $t_n$, whilst complicated, can be written down 
explicitly. They originate from the characteristic polynomial 
$\Phi(X) := \det(XI - M) = \sum_{k=0}^N c_k X^{N-k}$, with $c_0=1$. For any eigenvalue $\lambda_\nu$ we 
have $\Phi(\lambda_\nu)=0$ and thus it follows
\begin{equation}
\label{poleq}
t_{N+r}= \ -\left [\sum_{k=1}^{N} c_k t_{N+r-k}\right ].
\end{equation}
Newton's identities give the coefficients $c_k$ in terms of the $t_n, n\le k$ via the determinant
\begin{equation}
\label{newteq}
c_k =\frac{(-1)^k}{k!}
\left|\begin{array}{ccccc}
t_1 & 1   & 0 & \cdots & 0 \\
t_2 & t_1 & 2 & \cdots & 0 \\
\vdots & \vdots & \ddots &  \ddots & \vdots\\
t_{k-1} &  t_{k-2} & \cdots  & t_1& k-1\\
t_{k} & t_{k-1} & \cdots & t_2 & t_{1} \end{array}\right|.
\end{equation}
Therefore, using a combination of relations (\ref {poleq}) and (\ref{newteq}) one may write $G(\bt)$ 
explicitly in terms of the first $N$ traces $\bt$. Clearly $\Delta = G(\bt)^2$ is nothing but the discriminant 
of $\Phi(X)$ expressed as a function of $\bt$.

Using the transformation (\ref{PD transformation}) and the stationary distribution for the eigenvalues 
(\ref {spectrJPD}) we can obtain the JPDF for the traces in the three canonical ensembles
\begin{equation}\label{trJPD}
Q_{\beta} (\bt) = C^{(N)}_{\beta} G(\bt)^{\beta -1} \exp\left(-\frac{\beta}{2} t_2\right )\chi_{N}(\bt).
\end{equation}
$\chi_{N}(\bt)$ is an indicator function for the domain $\mathcal{T}\subset\mathbb{R}^N$ which is 
the support for $Q_{\beta} (\bt)$. In contrast to the spectrum, which is defined over the entire 
space $\mathbb{R}^N$, the trace parameters are restricted to the domain $\mathcal{T}$. This is 
because the traces are sums of powers of real variables, which must satisfy certain consistency 
relations: The inverse mapping $\bt \rightarrow \blambda$ should yield real spectra. For example, 
in 2 dimensions we have $2t_2-t_1^2 = (\lambda_2-\lambda_1)^2\ge 0$.  Hence, 
$  \mathcal{T}=\{(t_1,t_2)\in \mathbb{R}^2: 2t_2-t_1^2 \ge 0 \} $. In higher dimensions it 
becomes increasingly more difficult to write an explicit definition of $\mathcal{T}$, other 
than stating that it is the image of the mapping $\blambda \rightarrow \bt$. It should be emphasized, 
however, that $\mathcal{T}$ is independent of the ensemble under consideration - one may consider 
matrices with non-Gaussian elements, or even correlated elements, and $\mathcal{T}$ will remain the same.

We would also like to highlight that the GOE distribution takes a very simple form in this space, 
i.e. $Q_{1}(\bt)= C^{(N)}_{1}\exp\left(-\frac{1}{2 } t_2\right )$. At first sight it might seem 
strange that the JPDF for all the traces depends only on one parameter $t_2$; however, as alluded 
to above, one must pay very close attention to the domain of integration $\mathcal{T}$. This is 
exemplified in Section \ref{2 Dim example}, in which we calculate expectations values and marginal 
probabilities.

In the following section we shall derive the Fokker-Planck equation for $Q(\bt;s)$. Its stationary 
solution is known and given explicitly in (\ref{trJPD}). As will be shown below, by substituting 
this solution into the stationary Fokker-Planck equations we obtain two identities which are 
summarized in the following proposition.
\begin{prop}\label{Identities} For $n\ge0$ we have
\begin{eqnarray}
 2\,\sum_{m=1}^N m\,\frac{\p t_{n+m}}{\p t_m}
    & = & \sum_{i,\,j\ge0 \atop i+j=n} t_it_j \;+\; (n+1)\,t_n\, \label{Main ID 1} \\
 \frac{2}{G}\sum_{m=1}^N m\,t_{n+m}\,\frac{\p G}{\p t_m}
   & = & \sum_{i,\,j\ge0 \atop i+j=n}  t_it_j \;-\; (n+1)\,t_n\,. \label{Main ID 2}
   \end{eqnarray}
\end{prop}
\noindent As mentioned in the introduction, we are unaware of such identities arising before 
in RMT or any other context and a direct algebraic proof is given by D. Zagier in \ref{Zagier proof}.

\section {The Fokker-Planck equation}
\label{F-P equations}
The main reason for studying  Dyson's Brownian motion in the space of traces is that the 
Fokker-Planck equation for $Q(\bt;s)$ can be derived directly, avoiding the use of perturbation 
theory (\ref{Perturbation formula}). The expectation values of the components of $\bt$ due to 
an incremental changes in the matrices will be evaluated directly from the matrix elements statistics. 
Once $Q(\bt;s)$ has been computed, one can then transform back to the spectral representation in 
order to deduce the eigenvalue statistics though $P(\blambda,s)$.

We begin by expressing the change in the $n^{\rm th}$ trace via the change in the matrix $\delta M$, 
up to second order (since higher terms in $\delta M$ will be of orders $\delta s^2$ or greater 
after taking the expectation)
\begin{eqnarray}
\label{deltatn}
\delta t_n &=&\left[\tr((M+\delta M)^n) - \tr(M^n)\right]\nonumber \\
 &=& n\tr(M^{n-1}\delta M) + \frac{n}{2}\sum_{x=0}^{n-2}\tr(M^x \delta M M^{n-x-2} \delta M) + \ldots \; .
\end{eqnarray}
The simplest way to compute $\E(\delta t_n)$ and $\E(\delta t_n\delta t_m)$ is to invoke the 
invariance of the stochastic motion under unitary transformations. We are then free to write the 
initial matrix $M$ in a diagonal representation of eigenvalues, i.e. $M_{ij} = \lambda_i\delta_{ij}$. 
Using this and the expressions (\ref{Matrix elements general 1}) and (\ref{Matrix elements general 2}) we find
\begin{eqnarray}\label{First trace moment}
\E(\delta t_n) &=& -nt_n\delta s + \frac{n}{2}\sum_{x=0}^{n-2}\sum_{ijkl}\lambda_i^x\delta_{ij} \lambda_k^{n-x-2}\delta_{kl} \E(\delta M_{jk} \delta M_{li})
\nonumber
\\
&=& \left[-nt_n + \frac{n}{2}\sum_{x=0}^{n-2}t_x t_{n-2-x} + \frac{2-\beta}{\beta}\frac{n}{2}(n-1) t_{n-2}\right] \delta s,
\end{eqnarray}
where we have used that $\E(|\delta M_{ij}|^2) =  (1 + (2/\beta-1)\delta_{ij})\delta s$. In particular 
this means for $n=1$ and $2$ that we have $\E(\delta t_1) = -t_1\delta s$ and $\E(\delta t_2) = (-2t_2 + t_0^2 + (2/\beta-1)t_0)\delta s$.

For the second order moments, since again we need terms proportional to $\delta s$ and no more, we 
only require the first term in (\ref{deltatn}). Therefore, for $n,m =1,\ldots,N$, we get
\begin{eqnarray}
\label{Second trace moment}
\hspace{-20pt}
\E( \delta t_n \delta t_m) &=& nm\sum_{ijkl} \lambda_i^{n-1}\delta_{ij}\lambda_k^{m-1}\delta_{kl}   \E(\delta M_{ji}\delta M_{lk})\nonumber \\
 &=& nm\sum_{ik}\lambda_i^{n-1}\lambda_k^{m-1}\E(\delta M_{ii}\delta M_{kk}) = \frac{2nm}{\beta}t_{n+m-2}\delta s,
\end{eqnarray}
where we have used $\E(\delta M_{ii}\delta M_{kk}) = \frac{2}{\beta}\delta_{ik}\delta s$. Note that in 
the above equations, and in the following, one should remember that the independent parameters in the 
present theory are the components $\bt$ which consist of the first $N$ traces. Whenever there appears 
$t_x$ with $x>N$, it should be considered as a function of the independent parameters as explained in the previous section. Similarly, one must substitute $t_0 = N$.

We are now in a position to obtain our Fokker-Planck equation for determining the probability distribution 
$Q_{\beta}(\bt;s)$ of the traces. For simplicity we write (\ref{First trace moment})  and (\ref{Second trace moment}) 
in the form $R^{(\beta)}_n= \E(\delta t_n)/\delta s$ and $R^{(\beta)}_{nm}= \E(\delta t_n\delta t_m)/ \delta s$, 
so that (see for instance \cite{Wang})
\begin{equation}
\label{FP - traces}
\frac{\partial Q_{\beta}}{\partial s} = - \sum_n \frac{\partial (R^{(\beta)}_n Q_{\beta})} {\partial t_n} + \frac{1}{2} \sum_{n,m} \frac{\partial^2 (R^{(\beta)}_{nm} Q_{\beta})} {\partial t_n\partial t_m}.
\end{equation}
Just as $P_{\beta}(\blambda)$, given in (\ref{spectrJPD}), is the stationary solution to the Fokker-Planck 
equation (\ref{FP-spectral}) for the eigenvalues, so we would like to verify $Q_{\beta}(\bt)$, given 
in (\ref{trJPD}), is the stationary solution of (\ref{FP - traces}) above. For this to be the case, 
$Q_{\beta}(\bt)$ must therefore satisfy the following $N$ simultaneous equations
\begin{equation}
\label{statfokkerplanck}
   R^{(\beta)}_n Q_{\beta} = \frac{1}{2} \sum_{m} \frac{\partial (R^{(\beta)}_{nm} Q_{\beta})} {\partial t_m}, \ \ \ \ \forall\  1 \le n \le N.
\end{equation}
These will be discussed shortly for arbitrary matrix dimension $N$ but prior to this we outline, 
for illustrative purposes, the scenario for $N=2$.

\subsection{Example: $2\times 2$ Gaussian ensembles}\label{2 Dim example}
The $N=2$ case offers the particular advantage that the expressions (\ref{First trace moment}) 
and (\ref{Second trace moment}) do not contain traces larger than $t_N$ (i.e. $t_2$ in this case), 
which is not true for $N >2$. In order to satisfy (\ref{FP - traces}) $Q \equiv Q_{\beta}(t_1,t_2)$ 
must be a solution of the simultaneous equations (\ref{statfokkerplanck}), which in 2 dimensions are given by
\begin{eqnarray}
 0 & = & t_1 Q + \frac{1}{\beta}\left[\frac{(t_0\partial Q)}{\partial t_1} + 2\frac{\partial(t_1Q)}{\partial t_2}\right]
\nonumber \\
0 & = & \left(2t_2 - t_0^2 - \frac{(2-\beta)}{\beta} t_0 \right)Q + \frac{2}{\beta}\left[\frac{\partial (t_1Q)}{\partial t_1} + 2\frac{\partial(t_2Q)}{\partial t_2}\right].
  \nonumber
\end{eqnarray}
% 0&=& 2t_2 Q_2(t_1,t_2) -4Q_2(t_1,t_2)+ 2\p_{t_2}(t_2Q_2(t_1,t_2))
One may verify by substitution that the solution is, including the normalisation constant presented in (\ref{spectrJPD}),
\begin{equation}\label{22 Trace distribution}
Q_\beta(t_1,t_2) = \frac{1}{2}C^{(2)}_{\beta} \left(2t_2 - t_1^2\right)^{\frac{\beta-1}{2}}e^{-\frac{\beta t_2}{2}}.
\end{equation}
Written in terms of the eigenvalues, using $G(\bt)^2 = \left(2t_2 - t_1^2\right) = (\lambda_2 - \lambda_1)^2$, this yields
\[
P_\beta(\lambda_1,\lambda_2) = C^{(2)}_{\beta}|\lambda_2 - \lambda_1|Q(\lambda_1,\lambda_2) =
C^{(2)}_{\beta}|\lambda_2 - \lambda_1|^\beta e^{-\frac{\beta(\lambda_1^2 + \lambda_2^2)}{2}},
\]
which is the expected result for the JPDF.

From (\ref{22 Trace distribution}) we can immediately calculate the marginal probability distributions 
for the traces. Importantly, the limits of integration are defined by the domain $\mathcal{T}$. For 
2 dimensions this was outlined in Section \ref{definitions}
\begin{eqnarray}
q_\beta(t_1) & = & \int_{t_1^2/2}^{\infty}  d t_2 \; Q_\beta(t_1,t_2) = \frac{1}{2}C^{(2)}_{\beta} r_{\beta}e^{-\beta t_1^2/4} \\
q_\beta(t_2)  & = & \int_{-\sqrt{2t_2}}^{\sqrt{2t_2}} d t_1 \; Q_\beta(t_1,t_2) 
=  \frac{1}{2}C^{(2)}_{\beta} s_{\beta}t_2^{\beta/2}e^{-\beta t_2/2},
\end{eqnarray}
where $(C^{(2)}_{\beta})^{-1} = 4\sqrt{\pi}, \pi, 3\pi/8$,  $r_{\beta} = 2, \sqrt{\pi/2},3\sqrt{\pi}/8$ and 
$s_{\beta} = 2^{3/2},\pi,3\pi/2$ for $\beta = 1,2,4$ respectively. The expected value of $t_2$ is therefore 
$\langle t_2 \rangle = \int_0^{\infty} dt_2 \; t_2q_{\beta}(t_2) = 3,2,3/2$ in the three cases.

\subsection{Stationary solution}\label{Sec: Stationary solution}
Finding the stationary solution in $N$ dimensions requires solving the $N$ simultaneous equations given 
by (\ref{statfokkerplanck}). Therefore, substituting in the expressions (\ref{First trace moment}) and 
(\ref{Second trace moment}) we get for each $n$
\begin{equation}\label{identity}
\fl
\left(-nt_n  + \frac{n}{2}\sum_{x=0}^{n-2}t_x t_{n-2-x} + \frac{2-\beta}{\beta}\frac{n}{2}(n-1) t_{n-2}\right) Q_{\beta} 
= \frac{n}{\beta}\sum_{m=1}^N m\frac{\partial (t_{n+m -2}Q_{\beta})}{\partial t_m}.
\end{equation}
The derivative in the RHS can be expanded using the chain rule to obtain
\[
\fl
\frac{\partial (t_{n+m -2}G^{\beta-1}e^{-\beta t_2/2})}{\partial t_m} 
= \left(\frac{\partial t_{n+m -2}}{\partial t_m} + t_{n+m-2} \frac{(\beta -1)}{G}\frac{\partial G}{\partial t_n}
  - \frac{\beta}{2}t_{n+m -2}\delta_{2m}\right)Q_{\beta}.
\]
Therefore, after some algebra in which we divide through by a factor $nQ_{\beta}/(2\beta)$ and cancel the 
term involving $-nt_n$ on both sides, we arrive at the following relationship between the traces
\begin{equation}\label{Trace identity 0}
\fl
\beta \sum_{x=0}^{n-2}t_x t_{n-2-x} +  (2-\beta)(n-1) t_{n-2} = 2\sum_{m=1}^N m\left[
\frac{\partial t_{n+m -2}}{\partial t_m} + t_{n+m-2} \frac{(\beta -1)}{G}\frac{\partial G}{\partial t_m}\right].
\end{equation}
In the particular case $\beta=1$ there is no dependence on the Vandermonde determinant $G(\bt)$ and we get
\begin{equation}
\label{idgoe}
2\sum_{m=1}^N m \frac{\partial t_{n+m -2}}{\partial t_m} =  \sum_{x=0}^{n-2}t_x t_{n-2-x} +  (n-1) t_{n-2}.
 \end{equation}
Replacing $n-2$ by $n$ thus gives the identity (\ref{Main ID 1}). If we then rearrange (\ref{Trace identity 0}) 
in terms of $\beta$ we find
\[
\fl
 \beta\left (\sum_{x=0}^{n-2}t_x t_{n-2-x}-(n-1) t_{n-2}-
 2\sum_{m=1}^N m t_{n+m-2} \frac{1}{G}\frac{\partial G}{\partial t_m}\right)
 \]
 \begin{equation}\label{idbeta}
 = 2\left (\sum_{m=1}^N m\left[\frac{\partial t_{n+m -2}}{\partial t_m} - t_{n+m-2} \frac{1}{G}\frac{\partial G}{\partial t_m}\right]
- (n-1)t_{n-2}\right ).
\end{equation}
The above must be fulfilled simultaneously for both $\beta=2,4$, which only occurs if the expressions in large 
brackets on the two sides of (\ref {idbeta}) vanish. Therefore, by using the substitution (\ref{idgoe}) we 
arrive at the second identity (\ref{Main ID 2})
\[
 2\sum_{m=1}^N m t_{n+m-2} \frac{1}{G}\frac{\partial G}{\partial t_m} =
 \sum_{x=0}^{n-2}t_x t_{n-2-x} - (n-1) t_{n-2} \; ,
\]
where again we must replace $n-2$ by $n$. Since we know that the expression (\ref{trJPD}) must be our stationary 
solution the method above constitutes a proof of the identities (\ref{Main ID 1}) and (\ref{Main ID 2}). However,
 a direct proof of these is given by D. Zagier in \ref{Zagier proof}, which therefore implies that (\ref{trJPD})
 must be our stationary JPDF, without the need for any transformation of variables.

\subsection{The mean values $\langle t_n \rangle$}\label{Sec: Mean values}
Computations of expected values of any function of $\bt$ involve integrating over the domain $\chi_N(\bt)$, 
which is not explicitly defined for any $N>2$. However, one can use a simple heuristic reasoning in order to 
identify the mean values $\langle t_n \rangle$ as the coordinates of the vector $\bt$ for which the drift force 
(\ref{First trace moment}) vanishes, i.e.
\begin{equation}
\label{semicircle}
\langle t_n\rangle   = \frac{1}{2}\sum_{x=0}^{n-2}\langle t_x\rangle \langle t_{n-2-x}\rangle 
+ \frac{2-\beta}{2\beta}(n-1)\langle t_{n-2}\rangle .
\end{equation}
It is natural, and customary, to scale the matrices $M$ by $1/\sqrt{N}$ and the resulting traces by $1/N$, 
so that we may define $\tau_n = N^{-\frac{n}{2}-1} t_n $. Thus
\begin{equation}\label{Catalan}
\mean{\tau_n}   = \frac{1}{2}\sum_{x=0}^{n-2}\mean{\tau_x}\mean{\tau_{n-2-x}}  + \frac{2 - \beta}{2\beta N}(n-1) \mean{\tau_{n-2}}
\end{equation}
If we take $\beta = 2$, with initial conditions $\tau_0 =1$ and $\tau_1 =0$, then (\ref{Catalan}) implies 
that $\mean{\tau_{2k+1}} =0$ and $\mean{\tau_{2k}} =\frac{1}{2^k} C_k$ where $C_k$ are the Catalan numbers. 
This is the well known result obtained by computing the moments using the semi-circle spectral distribution 
function (see e.g. \cite {Anderson,Sinai,wigner}). For other $\beta$ the last term is of order 
$ \mathcal{O} (\frac{1}{N})$ smaller than the rest and thus its effect vanishes in the limit of large $N$. 
This is consistent with the fact that the spectral distribution of the three canonical ensembles converge 
to the semi-circle distribution for $N\rightarrow \infty$. Moreover for $N=2$, (\ref{semicircle}) returns 
$\langle t_2\rangle = 3,2, 3/2$ for $\beta = 1,2,4$ respectively, which is exactly the result obtained in 
Section \ref{2 Dim example}.

\section{Application to Bernoulli ensembles}
\label{Sec: Application}
Recently, the authors have used a discrete analogue of Dyson's Brownian motion model to investigate the 
spectral statistics of Bernoulli ensembles \cite{Joyner-2015}. Here we provide a brief illustration of 
how this can be adapted to the traces setting and discuss why this offers certain advantages. Our Bernoulli 
ensemble $\mathfrak{B}_N$ is given by the set of $N\times N$ symmetric matrices with 0 on the diagonal 
and off-diagonal entries chosen randomly and independently from the set $\{\pm a\}$ with equal probability 
(in the following we shall choose, without loss of generality, $a = 1/\sqrt{2}$ in order to match the 
variance of the GOE defined in Section \ref{introduction}). The spectral properties of $\mathfrak{B}_N$ 
were first analysed by E. Wigner in 1955, who showed the empirical spectral density converges to the 
semicircle distribution in the limit of large $N$ \cite{wigner}. Recent works have gone much further, 
establishing that local eigenvalue correlations do indeed converge to the corresponding Gaussian expressions 
as $N$ increases \cite{Tao-2011,Erdos-2011,Erdos-2013,Erdos-2012}.

In \cite{Joyner-2015} the random walk is defined on $\mathfrak{B}_N$ such that at each single time-step, 
one of the $d_N=\frac{1}{2}N(N-1)$ off-diagonal matrix entries $B_{pq}$ is chosen at random and its sign 
is flipped (together with $B_{qp}$). This leads to a change in the matrix $B$ of
\begin{equation}
\delta B^{pq} = -2 B_{pq}[\vec{p}\tvec{q} + \vec{q}\tvec{p}],
\end{equation}
where $\vec{p}$ is a vector whose elements are all zero but for $1$ in the position $p$, and $\tvec{p}$ 
is its transposed. This perturbation in turn induces a change in the eigenvalue $\lambda_{\mu}$ of
\begin{equation}\label{Perturbation formula 2}
\delta \lambda_{\mu} =\tvec{\mu} \delta B^{pq}\vec{\mu}+\sum_{\nu\ne\mu}\frac{|\tvec{\nu} \delta B^{pq}\vec{\mu}|^2}{\lambda_{\mu}-\lambda_{\nu}}
\ + \cdots \ ,
\end{equation}
in a similar manner to (\ref {Perturbation formula}). In order to construct the coefficients in the Fokker-Planck 
equation one has to average $\delta \lambda_{\mu}$  over the entire neighbourhood of matrices that can be reached 
in a single step. In particular, $\E(\tvec{\mu} \delta B\vec{\mu}) = -2\lambda_{\mu}/d_N$ and
\begin{equation}\label{EV second moment}
\fl
\E(|\tvec{\nu} \delta B \vec{\mu}|^2)=\frac{1}{d_N}\sum_{p<q} |\tvec{\nu} \delta B^{pq}\vec{\mu}|^2 
= \frac{2}{d_N}\left (1 + \delta_{\nu\mu} -2\sum_{p=1}^N \nu_p^2 \mu_p^2\right )\ .
\end{equation}
Here, in contrast to \cite{Joyner-2015}, there is an additional term $\sum_{p=1}^N \nu_p^2 \mu_p^2$ that cannot 
be written purely in terms of the eigenvalues, meaning the motion is not autonomous.

Collating the above expressions allows one to derive a Fokker-Planck equation which describes the motion 
of a suitable observable, up to an error that depends on $N$. This error comes from a combinations of factors 
such as higher moments $\E(\delta \lambda_{\mu}^k)$ and higher terms in the perturbation formula (\ref{Perturbation formula 2}). 
This is because, ultimately, our process in discrete and, unlike Dyson's Brownian motion, one cannot assume 
that the change of the matrix due to a single step can be made arbitrarily small. These errors, together 
with the correction to the second moment from the additional term in (\ref{EV second moment}), all depend 
on the eigenvectors and can only be assumed to become negligible in the large $N$ limit if they are sufficiently 
delocalised. For the present ensemble, it has been proved this is correct with high probability 
(see \cite{Tao-2011,Erdos-2011,Erdos-2013,Erdos-2012} and references therein) but for Bernoulli ensembles 
with correlated matrix entries there are no rigorous results thus far in this direction. Moreover, 
perturbation theory only converges when $|\tvec{\mu} \delta B\vec{\mu}|$ is small relative to 
$|\lambda_{\mu}-\lambda_{\mu\pm 1}|$. In ensembles such as random regular graphs, this is not the case, 
even though the eigenvectors are delocalised, due to the growth rate (or lack thereof) of the mean level spacing. 
These observations therefore motivate the search for another approach.

In complete analogy to Section \ref{F-P equations} we can also study the random walk in the space of traces. 
In fact we shall find it more amenable to use the rescaled traces $\tau_n = N^{-n/2-1}t_n$, as used in 
Section~\ref{Sec: Mean values}. In this basis all the variables are $\mathcal{O}(1)$ in $N$ and thus it 
becomes transparent as to which terms can be neglected. To facilitate this transition let us therefore scale 
the original matrices by $\bar{B} = B/\sqrt{N}$. Applying this to (\ref{deltatn}) we have
\begin{eqnarray}\label{Bernoulli trace expansion}
\delta \tau_n &=& \frac{1}{N}(\Tr(\bar{B} + \delta \bar{B})^n - \Tr(\bar{B}^n)) \nonumber \\
&=& \frac{1}{N}\left[n\Tr(\bar{B}^{n-1}\delta \bar{B}) 
+ \frac{n}{2}\sum_{x = 0}^{n-2} \Tr(\bar{B}^{x}\delta \bar{B} \bar{B}^{n-2-x}\delta \bar{B}) + \ldots\right].
\end{eqnarray}

Although we shall eventually seek to neglect those higher terms, as in Section \ref{F-P equations}, the 
whole expansion is finite for fixed $n$ and thus exact. It means this formalism offers a distinct advantage
over the perturbation formula (\ref{Perturbation formula 2}), which has no such guarantees. Moreover, 
in this way the change in the variables can be expressed directly in terms of the matrix elements, 
which is not the case for the eigenvalue representation, since it relies on the appearance of the eigenvectors.

Proceeding in a similar manner, the expected change of $\delta \tau_n$ in one time step may be 
calculated as follows\footnote{We use the convention that $B^n_{pq}$ denotes the $p,q$-th element 
of the matrix $B^n$ and $(B_{pq})^n$ is the matrix element $B_{pq}$ raised to the $n$-th power.}
\begin{eqnarray}
\label{eqappl1}
\fl
\E(\Tr(\bar{B}^{n-1}\delta \bar{B})) & = & -\frac{2}{d_N}\sum_{p<q}\bar{B}_{pq} \Tr\left (\bar{B}^{n-1}[\vec{p}\tvec{q}
 + \vec{q}\tvec{p}]\right) \nonumber \\
&= & -\frac{4}{d_N}\sum_{p<q}\bar{B}_{pq}\bar{B}^{n-1}_{pq} = -\frac{2}{d_N}N^{-n/2}t_n = -\frac{2}{d_N}N\tau_n
\end{eqnarray}
and
\begin{eqnarray}
\label{eqappl2}
\fl
\E\left (\Tr(\bar{B}^{x}\delta \bar{B} \bar{B}^{n-2-x}\delta \bar{B})\right )
&=& \frac{2}{d_N}\sum_{p<q} \bar{B}_{pq}^2 \Tr(B^{x}[\vec{p}\tvec{q}+\vec{q}\tvec{p}]\bar{B}^{n-2-x}[\vec{p}\tvec{q} + \vec{q}\tvec{p}])\nonumber \\
 &=&  \frac{2}{d_N}\frac{1}{N}\sum_{p \neq q} (\bar{B}^{x}_{pq} \bar{B}^{n-2-x}_{qp} + \bar{B}^x_{pp} \bar{B}^{n-2-x}_{qq})\nonumber \\
&=& \frac{2}{d_N}\left[\frac{1}{N}\sum_{p,q} (\bar{B}^{x}_{pq} \bar{B}^{n-2-x}_{qp} + \bar{B}^{x}_{pp} \bar{B}^{n-2-x}_{qq})
  - \frac{2}{N}\sum_p \bar{B}^{x}_{pp} \bar{B}^{n-2-x}_{pp}\right ] \nonumber \\
&=& \frac{2}{d_N}\left[\tau_{n-2} + N \tau_x \tau_{n-2-x} - 2\zeta(x,n-2-x)\right],
\end{eqnarray}
where
\[
\zeta(r,s) =\frac{1}{N}\sum_p \bar{B}^{r}_{pp} \bar{B}^{s}_{pp}.
\]
The most striking difference between (\ref{eqappl2}) and the Gaussian equivalent (\ref{First trace moment}) is the 
appearance of the term $\zeta(x,n-2-x)$, which cannot be expressed in terms of the variables $\bt$. Writing 
$\tau_s\tau_r -\zeta(r,s) = \frac{1}{N}\left(\sum_p \bar{B}^r_{pp}\left[\frac{1}{N}\sum_{q}\bar{B}^s_{qq} - \bar{B}^s_{pp}\right]\right)$ 
we see that $\zeta(r,s)$ is very close to $\tau_s\tau_r$ if the diagonal elements $\bar{B}^s_{pp}$ are close to their 
average over the whole diagonal $\sum_{q}\bar{B}^s_{qq}$. Using Wigner's combinatorial method of counting Dyck paths 
(see e.g. \cite{Anderson,wigner}) one can show that by averaging over $\BN$ we have for fixed $r$ and $s$ that 
$\langle \tau_r\tau_s - \zeta(r,s)\rangle_{\BN}$ tends to 0 as $N \to \infty$. Moreover, using the same technique 
one finds $\Var_{\BN}(\tau_r\tau_s - \zeta(r,s)) = \mathcal{O}(N^{-2})$. Hence with high probability $\zeta(r,s)$ 
is $\mathcal{O}(1)$. This shows that (\ref{eqappl2}) is dominated by the term $N\tau_x\tau_{n-2-x}$. In addition 
we can also estimate those higher terms in the expectation $\E(\delta \tau_n)$ coming from the expansion 
(\ref{Bernoulli trace expansion}). For example, we have $\E(\Tr(\delta \bar{B}^3 \bar{B}^{n-3})) 
= N^{-n/2}\E(\Tr(\delta B^3 B^{n-3})) = 4N^{-n/2}\E(\Tr(\delta B B^{n-3})) = -8N^{-n/2}t_{n-2}/d_N = -8\tau_{n-2}/d_N$. 
This again is of an order in $N$ less than the dominant term in (\ref{eqappl2}). Therefore, in the large $N$ limit 
we find that $\E(\delta \tau_n)/\delta s$ (taking $\delta s = 2/d_N$) tends to the expression (\ref{First trace moment}) calculated for the GOE.

Similarly, for the second moment we find
\begin{equation}\label{eqappl3}
\E(\delta \tau_n\delta \tau_m)= \frac{2}{d_N}\frac{2nm}{N^2}\left(\tau_{n+m-2}-\zeta(n-1,m-1)\right ) + \ldots \; .
\end{equation}
The difference in comparison to the first moment is that, by the arguments above, the additional term $\zeta(n-1,m-1)$ is of the same order in $N$ as the supposed leading term. This is also in contrast to the outcome for the second 
order term in the eigenvalue representation (\ref{EV second moment}), where the effect of removing the matrix 
diagonal leaves only a $1/N$ correction. Nevertheless we present arguments that allow for it to be neglected. 
Let us continue by inserting the expressions (\ref{eqappl1}), (\ref{eqappl2}) and (\ref{eqappl3}) into the 
appropriately scaled version of the $n$ simultaneous equations (\ref{statfokkerplanck}), which determine the 
stationary solution $Q$ (the method for calculating the error terms in the analogous eigenvalue representation 
approach is discussed at length in \cite{Joyner-2015} and thus we refrain from details here). Therefore, for 
large $N$, the stationary solution $Q$ for $\mathfrak{B}_N$ approximately satisfies
\[
\fl
\left[-\tau_n + \sum_{x=0}^{n-2}\left(\tau_x\tau_{n-2-x} + \frac{\tau_{n-2}}{N}\right)\right]Q
 = \sum_m \frac{2m}{N^2} \frac{\partial}{\partial \tau_m}\left\{(\tau_{n+m-2} - \zeta(n-1,m-1) )Q\right\}.
\]
To estimate the contribution of $\zeta(n-1,m-1)$ we replace the exact value with its 
mean, i.e. $N^{-1}\sum_p \bar{B}^{n-1}_{pp} \bar{B}^{m-1}_{pp} \approx \tau_{n-1}\tau_{m-1}$. For all 
matrices $B \in \mathfrak{B}_N$ we have $\tau_1=0$ and $\tau_2 = N(N-1)/N^2 = 1 - 1/N$, meaning our space 
of variables is reduced to $\tau_n$ for $n=3,\ldots N$. Assuming then, that in all the remaining directions 
our JPDF $Q$ is constant (as is the case in the GOE expression (\ref{trJPD})) we find for $n \geq 3$
\[
\sum_m \frac{2m}{N^2} \frac{\partial}{\partial \tau_m}\left(\tau_{n-1}\tau_{m-1}Q\right) = \frac{2}{N^2}(n-1)\tau_{n-2}Q,
\]
where we have used that $\partial \tau_{m-1}/\partial \tau_m = 0$ and $\partial \tau_{n-1}/\partial \tau_m = \delta_{m,n-1}$ 
for all $n,m \leq N$. This results in a term which is of order $1/N$ less than the corresponding term on the LHS 
and a full order $1/N^2$ less than the leading term.

 %If $n=m$ (and $n$ is even) we get asymptotically for the ratio of the means of each term $C_{(n-1)/2}^2/C_{(n+m-2)/2} \sim ((2n -2)/(n-1)^2)^{3/2} \sim n^{-3/2}$. Hence as $n$ becomes large those cross-correlations for $n \approx m$ are governed more and more by the leading term.
 
\section{Discussion}\label{Sec: Discussion}
The efforts invested in developing the formalism presented above were motivated by our initial observations 
regarding random regular graphs. Dyson's original model could not be transcribed to this matrix ensemble as 
the perturbation formula is effectively useless (a consequence of small separation between eigenvalues) in 
this context (see \cite{Joyner-2015b}). Here we offer a method which does away with the requirement of the 
perturbation formula and therefore offers a potential method for circumventing such problems. We have demonstrated 
this method in the standard Gaussian setting and also illustrated how this can be used for Bernoulli matrices. 
The former case leads immediately to two previously unseen identities regarding symmetric functions, which are 
proved directly below. Finally we also note the relation with those studies 
\cite{Sinai,Essen,Guionnet,Diaconis-1994,Diaconis-2004} regarding the distributions of traces. Except for 
\cite{Guionnet}, these works did not consider any dynamical aspects and so what we have outlined here may 
offer alternative ways for studying traces distributions. For instance one should be able to apply the same 
techniques to the circular ensembles.

\section*{Acknowledgements}
US acknowledges the Institut Henri Poincar\'{e} for the hospitality extended when the manuscript was put in 
its final form. CHJ thanks the Isaac Newton Institute for their hospitality during the writing of this article 
and acknowledges the financial support of both the Feinberg Graduate School and Leverhulme Trust (grant 
number ECF-2014-448). US and CHJ would also like to extend their gratitude to D. Zagier for providing a very 
nice proof of the identities in Section \ref{definitions} and writing the following appendix. We also thank 
P. Forrester for bringing to our attention the reference of Bakry and Zani.

\appendix

\section{Proof of Proposition \ref{Identities} by Don Zagier}\label{Zagier proof}

Following the notation of the paper, we let $\l_\a$ ($\a=1,\dots,N$) be independent variables 
and let $c_i$ ($0\le i\le N$), $t_n$~($n=0,1,\dots$) and $\D$ (discriminant) be the elements
of the algebra $S =\C[\l_1,\dots,\l_N]^{\frak S_N}$ of symmetric polynomials in the~$\l_\a$ defined by
\begin{eqnarray*}
 &&\Phi(X) \;:=\; \prod_{\a=1}^N(X-\l_\a) \= \sum_{i=0}^N c_iX^i\,,  \\
&& t_n\=\sum_{\a=1}^N\l_\a^n\,, \qquad \D \= {\rm disc}(\Phi) \;= \prod_{1\le\a<\b\le N}(\l_\a-\l_\b)^2 \;.
 \end{eqnarray*}
For $n<0$ we set $t_n=0$.  We have $c_N=1$ and $t_0=N$, while both $(c_1,\dots,c_N)$ and $(t_1,\dots,t_N)$
generate the algebra~$S$.  In particular, if we take the latter as coordinates on~$S$, then we can ask for the 
values of $\p t_n/\p t_m$ and $\p\D/\p t_m$ for $n\ge0$ and  $1\le m\le N$.  (Of course the former is $\d_{nm}$
for $0\le n\le N$, so it is only interesting if~$n>N$.)

The identities (\ref{Main ID 1}) and (\ref{Main ID 2}) were proved in the body of this paper using an indirect 
proof coming from random matrix theory.  Here we give a purely algebraic verification of both of these 
identities, and some small generalizations. For the reader's convenience we repeat these identities 
here, expressing the second one in terms of the polynomial invariant~$\D$ rather than its square-root~$G$.
\begin{prop*}For $n\ge0$ we have
\begin{eqnarray}
 &&\quad\;\,\, 2\,\sum_{m=1}^N m\,\frac{\p t_{n+m}}{\p t_m}
    \;\= \sum_{i,\,j\ge0 \atop i+j=n} t_it_j \;+\; (n+1)\,t_n\;,  \label{App: Main ID 1} \\
 && \,\frac1\D\,\sum_{m=1}^N m\,t_{n+m}\,\frac{\p\D}{\p t_m}
  \= \sum_{i,\,j\ge0 \atop i+j=n}  t_it_j \;-\; (n+1)\,t_n\;. \label{App: Main ID 2} 
   \end{eqnarray}
\end{prop*}
We use that the logarithmic derivative of $\Phi(X)$ is a generating series for the~$t_n$, i.e.,
\[
T(X) \;:= \frac{\Phi'(X)}{\Phi(X)} \= \sum_{\a=1}^N\frac1{X-\l_\a} \= \sum_{n=0}^\infty\frac{t_n}{X^{n+1}}\,,
\]
where the last expression can be taken either as a formal power series in $S[[1/X]]$
or as a holomorphic function in the annulus $|X|>\max_\a|\l_\a|$ if the~$\l_\a$ are complex numbers.
Dividing (\ref{App: Main ID 1}) and (\ref{App: Main ID 2}) by $X^{n+2}$ and summing over $n\ge-m$ (or equivalently~$n\ge0$, 
since $\p t_{n+m}/\p t_m$ vanishes for $-m\le n<0$), we can rewrite these two identities as
\begin{equation}\label{App: Main ID 1b}
 \quad 2\,\sum_{m=1}^Nm\,\frac{\p T(X)}{\p t_m}\,X^{m-1} \= T(X)^2 \m T'(X)
\end{equation}
and
\begin{equation}\label{App: Main ID 2b}
\frac{T(X)}\D\,\sum_{m=1}^Nm\,\frac{\p\D}{\p t_m}\,X^{m-1} \= T(X)^2 \+ T'(X)\,.
\end{equation}
For the proof, we define polynomials $\Phi_\a(X)$ and coefficients $c_{\a,n}$ for $1\le\a\le N$ and $0\le n\le N-1$ by
\[
\Phi_\a(X)\=\prod_{\b\ne\a}\frac{X-\l_\b}{\l_\a-\l_\b} \= \frac1{\Phi'(\l_\a)}\,\frac{\Phi(X)}{X-\l_\a}
  \= \sum_{n=0}^{N-1}c_{\a,n}\,X^n\,.
\]
Then $\Phi_\a(\l_\b)=\d_{\a\b}$, so $(c_{\a,n})$ is the inverse of the Vandermonde matrix~$(\l_\a^n)_{n,\a}$. On
the other hand, we have $\frac{1}{m}\,\frac{\p t_m}{\p\l_\a}=\l_\a^{m-1}$, so $c_{\a,m-1}=m\,\frac{\p\l_\a}{\p t_m}$
for $1\le m\le N$. Hence
\begin{equation} \label{App: Relation 1}
m\,\frac{\p T(X)}{\p t_m} \= \sum_{\a=1}^Nc_{\a,m-1}\,\frac{\p T(X)}{\p\l_\a}
  \= \sum_{\a=1}^N\,\frac{c_{\a,m-1}}{(X-\l_\a)^2}\,,
\end{equation}
so each term $\p T(X)/\p t_m$ is a rational function of the form $P_m(X)/\Phi(X)^2$ where~$P_m(X)$ is a polynomial
of degree~$\le2n-2$. Multiplying (\ref{App: Relation 1}) by $X^{m-1}$ and summing over~$m=1,\dots,N$ gives
\begin{eqnarray} \fl
 \qquad\quad\sum_{m=1}^N\,m\, \frac{\p T(X)}{\p t_m}\,X^{m-1} &&\= \sum_{\a=1}^N\,\frac{\Phi_\a(X)}{(X-\l_\a)^2} \nonumber \\
 &&\= \Phi(X)\,\sum_{a=1}^N\frac{1}{\Phi'(\l_\a)}\,\frac1{(X-\l_\a)^3}   \nonumber \\
 &&\= \Phi(X) \,\sum_{a=1}^N \Res_{z=\l_\a}\biggl(\frac1{\Phi(z)}\,\frac{dz}{(X-z)^3}\biggr) \nonumber \\
 &&\= \Phi(X)\;\Res_{z=X}\biggl(\frac{dz}{(z-X)^3\,\Phi(z)}\biggr)  \nonumber \\  
 &&\= \frac{\Phi(X)}2\,\frac{d^2}{dX^2}\frac1{\Phi(X)} \;\,=\;\, -\,\frac12\,\frac{\Phi''(X)}{\Phi(X)}\+\frac{\Phi'(X)^2}{\Phi(X)^2} \nonumber \\
 &&\= -\,\frac{T'(X)}2\+\frac{T(X)^2}2\,,\nonumber
 \end{eqnarray}
where in the fourth line we have used the residue theorem.  This prove the first identity~\eqref{App: Main ID 1b}. 
The calculation for~\eqref{App: Main ID 2b} is similar. We have
\[
\fl
\frac m{2\D}\,\frac{\p\D}{\p t_m} \= \frac12\,\sum_{\a=1}^Nc_{\a,m-1}\,\frac{\p\log\D}{\p\l_\a}
\= \sum_{\a=1}^Nc_{\a,m-1}\sum_{\b\ne\a}\frac1{\l_\a-\l_\b} \= \sum_{\a=1}^Nc_{\a,m-1}\,\frac{\Phi'_\a(\l_\a)}{\Phi_\a(\l_\a)}\;.
\]
Substituting into this the identity
\begin{eqnarray} \fl
\quad\qquad\frac{\Phi'_\a(\l_\a)}{\Phi_\a(\l_\a)} &&\= \biggl(\frac{\Phi'(t)}{\Phi(t)}\m\frac1{t-\l_\a}\biggr)\biggr|_{t=\l_\a}
  =\; \biggl(\frac{\Phi'(\l_\a+\e)}{\Phi(\l_\a+\e)}\m\frac1\e\biggr)\biggr|_{\e=0} \nonumber \\
&& \= \biggl(\frac{\Phi'(\l_\a)+\Phi''(\l_\a)\,\e\+\cdots}{\Phi'(\l_\a)\,\e
      \+\frac12\,\Phi''(\l_\a)\,\e^2\+\cdots}\m\frac1\e\biggr)\biggr|_{\e=0} 
 \= \frac12\,\frac{\Phi''(\l_\a)}{\Phi'(\l_\a)} \,,\nonumber
 \end{eqnarray}
multiplying by $X^{m-1}$ and summing over~$m$, we obtain the second identity~\eqref{App: Main ID 2b}\,:
\begin{eqnarray} \fl
\qquad\frac1\D\, \sum_{m=1}^Nm\,\frac{\p\D}{\p t_m}\,X^{m-1} &&\= \sum_{\a=1}^N\frac{\Phi''(\l_\a)}{\Phi'(\l_\a)}\,\Phi_\a(X)
   = \sum_{\a=1}^N\frac{\Phi''(\l_\a)}{\Phi'(\l_\a)^2}\,\frac{\Phi(X)}{X-\l_\a} \nonumber \\
 &&\= \Phi(X)\, \sum_{\a=1}^N\Res_{z=\l_\a}\biggl(\frac{\Phi''(z)}{\Phi'(z)}\,\frac{dz}{(X-z)\Phi(z)}\biggr) \nonumber \\
 &&\= \Phi(X)\, \Res_{z=X}\biggl(\frac{\Phi''(z)\,dz}{(z-X)\Phi(z)\Phi'(z)}\biggr) \nonumber \\
 &&\= \frac{\Phi''(X)}{\Phi'(X)} \= T(X) \+ \frac{T'(X)}{T(X)}\;.  \qquad\qquad\qquad\square \nonumber
 \end{eqnarray}

\medskip
We mention that one can use the same method of calculation to obtain other identities of this type.  For instance,
\begin{eqnarray} \fl
 \qquad \sum_{m=1}^N m (m-1)  \frac{\p T(X)}{\p t_m}X^{m-2} &&\= \sum_{\a=1}^N\,\frac{\Phi'_\a(X)}{(X-\l_\a)^2} \nonumber \\
&&\= \sum_{a=1}^N\frac1{\Phi_\a'(\l_\a)}\,\biggl(\frac{\Phi'(X)}{(X-\l_\a)^3} \m \frac{\Phi(X)}{(X-\l_\a)^4}\biggr)  \nonumber \\
&&\= \Res_{z=X}\biggl[\Bigl(\frac{\Phi'(X)}{(z-X)^3} \+ \frac{\Phi(X)}{(z-X)^4}\Bigr)\frac{dz}{\Phi(z)}\biggr] \nonumber \\
&&\=  \frac{\Phi'(X)}2\,\Bigl(\frac1{\Phi(X)}\Bigr)'' \+ \frac{\Phi(X)}6\,\Bigl(\frac1{\Phi(X)}\Bigr)''' \nonumber \\
&&\=  \frac13\, T(X)^3 \m \frac16\,T''(X) \nonumber
  \end{eqnarray}
and hence, in analogy with~\eqref{App: Main ID 1},
\[ 3\,\sum_{m=1}^N  m(m-1)\,\frac{\p t_{n+m}}{\p t_m}
\= \sum_{i,\,j,\,k\ge0 :\atop i+j+k=n} t_it_jt_k \m \frac{(n+1)(n+2)}2\,t_n\;. \]
Identities with polynomials of higher degree in~$m$ on the left could be proved in the same way.


\section*{References}
 \begin{thebibliography}{9}
\bibitem{Dyson1}
F. J. Dyson, \emph{A Brownian-motion model for the eigenvalues of a random matrix}, J. Math. Phys. {\bf 3} 1191-1198 (1962).

\bibitem{mehtabook}
M. L. Mehta, \emph{Random Matrices}, Third Edition, 142, Pure and Applied Mathematics (Elsevier/Academic Press, Amsterdam, 2004).

\bibitem{forrester}
P. Forrester, \emph{Log-Gases and Random Matrices}, London Mathematical Society Monographs \textbf{34}, (Princeton University Press, 2010).

\bibitem{Anderson} G. W. Anderson, A. Guionnet, and O. Zeitouni, \emph{An Introduction to Random Matrices}, Cambridge Studies in Advanced Mathematics {\bf 118} (Cambridge University Press, 2009).

\bibitem{Guhr} T. Guhr, A. M\"{u}ller-Groeling, and H. A. Weidenm\"{u}ller, \emph{Random matrix theories in quantum physics: Common concepts}, Phys. Rep. 299, 189 (1998).
  
\bibitem{Akemann}  G. Akemann, J. Baik, and P. Di Francesco (Ed.), \emph{The Oxford Handbook of Random Matrix Theory}, (Oxford University Press, 2011).

\bibitem{Porter} C. E. Porter, \emph{Statistical Theories of Spectra: Fluctuations} (Academic Press, New York, 1965).

\bibitem{Haakebook} F. Haake, \emph{Quantum Signatures of Chaos}, Springer Series in Synergetics, Third Edition, (Springer-Verlag, Berlin 2010).

\bibitem{Sinai}
Y. Sinai and A. Soshnikov, \emph{Central limit theorem for traces of large random matrices with independent matrix elements}, Bol. Soc. Bras. {\bf 29} 1-24 (1998).

\bibitem{Essen}
F. Haake, M. Ku\'{s}, H.-J. Sommers, H. Schomerus and K. \.{Z}yczkowski, \emph {Secular determinants of random unitary matrices} J. Phys. A: Math. Gen. 29 (1996) 3641–3658.

\bibitem{Guionnet}
A Guionnet, \emph{Uses of free probability in random matrix theory}, XVIth International Congress on Mathematical Physics (2010) pp. 106-122.

\bibitem{Diaconis-1994}
P. Diaconis and M. Shahshahani, \emph{On the eigenvalues of random matrices}, J. Appl. Probab. 31A (1994), 49?62.

\bibitem{Diaconis-2004}
P. Diaconis and A. Gamburd, \emph{Random matrices, magic squares and matching polynomials}, Electron. J. Combin. 11 (2004/06), no. 2, Research Paper 2, 26 pp.

\bibitem{Bakry-2014}
D. Bakry and M. Zani, \emph{Dyson processes associated with associative algebras: The Clifford case}, in \emph{Geometric Aspects of Functional Analysis} (eds. B. Klartag, E. Milman), Lecture Notes in Mathematics Volume 2116, pp 1-37 (Springer International Publishing Switzerland 2014).

\bibitem{Dunne-1993}
G. V. Dunne, \emph{Slater decomposition of Laughlin states}, Int. Journ. Mod. Phys. N 7 (28), 4783 (1993).

\bibitem{Vivo-2008}
P. Vivo and S. N. Majumdar, \emph{On invariant $2 \times 2$ $\beta$-ensembles of random matrices}, Phys. A. 387 (2008), no. 19-20, 4839-4855.

\bibitem{Joyner-2015}
C. H. Joyner and U. Smilansky, \emph{Spectral statistics of Bernoulli matrix ensembles - a random walk approach (I)}, preprint (2015), \url{http://arxiv.org/abs/1501.04907}.

\bibitem{wigner} Eugene P. Wigner, \emph{Characteristic vectors of bordered matrices with infinite dimensions}, The Annals of Mathematics, 2nd Ser. {\bf 62}, 548-564 (1955).

\bibitem{Tao-2011}
T. Tao and V. Vu, \emph{Random matrices: Universality of the local eigenvalue statistics}, Acta Math. 206 (2011), 127-204.

\bibitem{Erdos-2011}
L. Erd\H{o}s, H.-T. Yau and J. Yin, \emph{Universality for generalized Wigner matrices with Bernoulli distribution}, J. of Combinatorics, 1 (2011), no. 2, 15–85

\bibitem{Erdos-2013}
L. Erd\H{o}s, A. Knowles, H.-T. Yau and J. Yin, \emph{Spectral statistics of Erd\H{o}s-R\'enyi graphs I: Local semicircle law.}, Ann. Probab. (2013) 41, no. 3B, 2279-2375 .

\bibitem{Erdos-2012}
L. Erd\H{o}s, A. Knowles, H.-T. Yau and J. Yin, \emph{Spectral statistics of Erd\H{o}s-R\'enyi graphs II: Eigenvalue spacing and the extreme eigenvalues}, Comm. Math. Phys. 314 (2012), no. 3. 587-640.

\bibitem{Wang}
M. C. Wang and G. E. Uhlenbeck, \emph{On the theory of the Brownian motion II}. Revs. Mod. Phys. {\bf 17} 323342 (1945).

\bibitem{Joyner-2015b}
C. H. Joyner and U. Smilansky, \emph{Spectral statistics of Bernoulli matrix ensembles - a random walk approach (II)}, in preparation.


\end{thebibliography}
\end{document}